\documentclass[fleqn,usenatbib]{mnras}

\usepackage{newtxtext,newtxmath}
\usepackage{graphicx}
\usepackage[T1]{fontenc}
\usepackage{ae,aecompl}
\usepackage{graphicx}	
\usepackage{amsmath}	
\usepackage{amssymb}

\title[Timing noise \& spin frequency second derivatives]{Correlated timing noise and high precision pulsar timing: Measuring frequency second derivatives as an example}

\author[X. J. Liu et al.]{
X. J. Liu,$^{1}$\thanks{E-mail: xiao-jin.liu@postgrad.manchester.ac.uk}
M. J. Keith,$^{1}$
C. G. Bassa,$^{2}$
B. W. Stappers$^{1}$
\\
$^{1}$Jodrell Bank Centre for Astrophysics, School of Physics and
Astronomy, The University of Manchester, Manchester M13 9PL, UK\\
$^{2}$ASTRON, the Netherlands Institute for Radio Astronomy, Oude Hoogeveensedijk 4, 7991 PD  Dwingeloo, The Netherlands\\
}

\date{Accepted XXX. Received YYY; in original form ZZZ}

\pubyear{2019}

\begin{document}
\label{firstpage}
\pagerange{\pageref{firstpage}--\pageref{lastpage}}
\maketitle

\begin{abstract}

We investigate the impact of noise processes on high-precision pulsar timing. Our analysis focuses on the measurability of the second spin frequency derivative $\ddot{\nu}$. This $\ddot{\nu}$ can be induced by several factors including the radial velocity of a pulsar. We use Bayesian methods to model the pulsar times-of-arrival in the presence of red timing noise and dispersion measure variations, modelling the noise processes as power laws. Using simulated times-of-arrival that both include red noise, dispersion measure variations and non-zero $\ddot{\nu}$ values, we find that we are able to recover the injected $\ddot{\nu}$, even when the noise model used to inject and recover the input parameters are different. Using simulations, we show that the measurement uncertainty on $\ddot{\nu}$ decreases with the timing baseline $T$ as
$T^\gamma$, where $\gamma=-7/2+\alpha/2$ for power law noise models with shallow power law indices $\alpha$ ($0<\alpha<4$). For steep power law indices ($\alpha>8$), the measurement uncertainty reduces
with $T^{-1/2}$. We applied this method to times-of-arrival from the European Pulsar Timing Array and the Parkes Pulsar Timing Array and determined $\ddot{\nu}$ probability density functions for 49 millisecond pulsars. We find a statistically significant $\ddot{\nu}$ value for PSR\,B1937+21 and consider possible options for its origin. Significant (95 per cent C.L.) values for $\ddot{\nu}$ are also measured for PSRs\,J0621+1002 and J1022+1001, thus future studies should consider including it in their ephemerides. For binary pulsars with small orbital eccentricities, like PSR\,J1909$-$3744, extended ELL1 models should be used to overcome computational issues. The impacts of our results on the detection of gravitational waves are also discussed. 

\end{abstract}


\section{Introduction}
Rapidly rotating millisecond pulsars (MSPs) are recognised as excellent celestial clocks. Pulsar timing is the technique of measuring the times of arrival (TOAs) of pulses and using them to form a timing model, which accounts for rotational, astrometric and, if applicable, orbital parameters of a pulsar and models the observed change in arrival time. 

Pulsar timing has become an important tool in pulsar research and has been used for many applications, such as measuring or constraining gravitational radiation of close binary systems \citep{wt81,wh16}, testing theories of gravity \citep{agh+18}, studying the ephemeris of, and identifying potentially unknown objects in, our Solar System \citep{chm+10, glc18, abb1+18}, and investigating interstellar plasma \citep{yhc+07,kcs+13}, only to name a few.    

One of the key applications of pulsar timing is detecting nanohertz gravitational waves of various origins \citep{jb03,wl03,zhao11,mcc17}, by observing many MSPs comprehensively \citep{fb90}. These MSPs form a pulsar timing array (PTA). Several PTAs have been set up to target nanohertz gravitational waves, including the European Pulsar Timing Array (EPTA, \citealt{dcl+16}), the Parkes Pulsar Timing Array (PPTA, \citealt{rhc+16}), the North American Nanohertz Observatory for Gravitational Waves (NANOGrav, \citealt{abb1+18}) and the synergetic project of the International Pulsar Timing Array (IPTA, \citealt{vlh+16,lsc+16}).

These PTAs are working on many different approaches, from improving hardware through to improving techniques to improve the timing precision and thus increase the sensitivity to gravitational waves. One of these investigations was presented in \citet{lbs18}, in which we considered the impact of unmodelled effects of pulsar motion in our Galaxy on high precision timing. We proposed that the radial velocity of an MSP may contribute significantly to the spin frequency second derivative, $\ddot{\nu}$.
Depending on the properties of an MSP, including the radial velocity, $\ddot{\nu}$ could range from $10^{-31}$\,s$^{-3}$ to $10^{-28}$\,s$^{-3}$. As such, this term may induce noticeable timing residuals and affect the precision of PTAs in the long run \citep[also see][]{bp93,van03a}. Assuming no  correlations in the TOA residuals (thus only white timing noise in the timing data) and even cadence, \citet{lbs18} showed that the measurement error of $\ddot{\nu}$ decreases as $T^{-7/2}$, where $T$ is the timing baseline. Furthermore, $\ddot{\nu}$ larger than $2\times 10^{-29}$\,s$^{-3}$ could be detected with good confidence for the MSPs in the three PTAs \citep{lbs18}. As a by-product, this detection may measure the radial velocity, which is important for studying the Galactic orbit \citep{fbw+11,avk+12,bjs+16} and the formation of MSPs \citep{tb96}. 

The assumption of no correlation in the timing residuals is usually not valid, as two kinds of correlations: the dispersion measure (DM) variations \citep{yhc+07} and observing frequency independent correlated noise, which we term ``red noise" throughout this paper, have been observed in many MSPs \citep[e.g.][]{lsc+16}. The variations in DM are caused by the change of plasma density along the line of sight due to the turbulent motion of the interstellar medium and the relative motion of the interstellar medium and the pulsar \citep{ric77,fc90,ars95}. The physical origin of red noise is not well understood, although theories relating to rotational instability, magnetospheric changes and unmodelled pulsar companions have been proposed, see \cite{cll+16} for a summary. If they are not properly dealt with, these correlations can cause serious biases on the estimation of model parameters and their errors \citep{chc+11}. Parameter and uncertainty estimation thus need to incorporate analyses on DM variation and red noise.

In this paper, we aim to measure $\ddot{\nu}$ and estimate its measurement error by incorporating white noise and processes of correlation from DM variations and red noise. The structure of this paper is as follows. We introduce the method we used in Section\,\ref{fitting}. We then describe the noise models we applied to test the method in Section\,\ref{simulations}. The timing data we used are introduced in Section\,\ref{data}. We present and discuss the results in Section\,\ref{results}, and finally summarize our conclusions in Section\,\ref{conclusions}.

\section{Method}
\label{fitting}

To solve for red and white noise parameters at the same time as the pulsar parameters, and to provide a robust method to estimate $\ddot\nu$ in the presence of correlated noise, we use a Bayesian method \citep{lah+14}. We make use of the Enhanced Numerical Toolbox Enabling a Robust PulsaR Inference SuitE\footnote{\url{https://enterprise.readthedocs.io/en/latest/index.html}} (\textsc{enterprise}) to construct the models for DM variations and red noise, combined with \textsc{tempo2}\footnote{\url{https://bitbucket.org/psrsoft/tempo2}}\citep{hem06,ehm06} to determine the parameters of pulsar timing models. Sampling of the parameter space is done using parallel tempering Markov-chain Monte Carlo using the publicly available python implementation \textsc{PTMCMCSampler}\footnote{\url{http://jellis18.github.io/PTMCMCSampler/index.html}}\citep{evh17}.

We model the TOAs using the pulsar rotational spin, astrometric and orbital parameters of the pulsar specified in the pulsar timing models provided with the data set \citep{dcl+16,rhc+16}, plus a noise model generated by \textsc{enterprise}. For efficiency of computation, \textsc{enterprise} uses the linearised model of \textsc{tempo2} to analytically marginalise over the pulsar timing parameters, except for $\ddot\nu$. The noise model is a combination of white noise parameters for each instrument, a power law red noise model and a power law DM variation model. These noise models are described in detail in \citet{lah+14}. We assumed uniform flat priors for all noise and DM variation parameters, with a range covering typical values for millisecond pulsars. 

For each pulsar and data set we used \textsc{PTMCMCSampler} to estimate the posterior distribution of the model parameters by computing $10^6$ iterations of the Markov chain. After discarding $25\%$ as a burn-in and thinning the chain by a factor of 10, we had $7.5\times10^4$ samples of the posterior distribution. In this paper we consider only the posterior distribution of $\ddot{\nu}$, marginalising over the remaining parameters. As there is a covariance between $\ddot{\nu}$ and the red noise parameters, the posterior distribution of $\ddot{\nu}$ typically has a wider tail than a Gaussian distribution. Therefore we present the results as the $95$ per cent confidence interval in addition to the  mean and standard deviation.

\section{Simulations}
\label{simulations}
In order to assess the effectiveness of our method for recovering a $\ddot{\nu}$ in the timing data, we first apply the method to the data in which a $\ddot{\nu}$ of known value, $\ddot{\nu}_{\rm in}$, was added to a simulation of 1000 realisations of TOAs from PSR J0437$-$4715. Each realisation of the simulation consists of 512 TOAs chosen to have zero residual relative to the timing ephemeris, roughly evenly spaced across a time-span of 15 years, with TOAs at alternating observing frequencies of 1432 MHz and 610 MHz. Uniform Gaussian white noise with zero mean and $\sigma=30$\,ns was added to each TOA.
To these residuals we then add DM variations and red noise as described in the following sections. 

\subsection{The frequency second derivative}
\begin{figure}
 \centering
 \includegraphics[width=8.6cm]{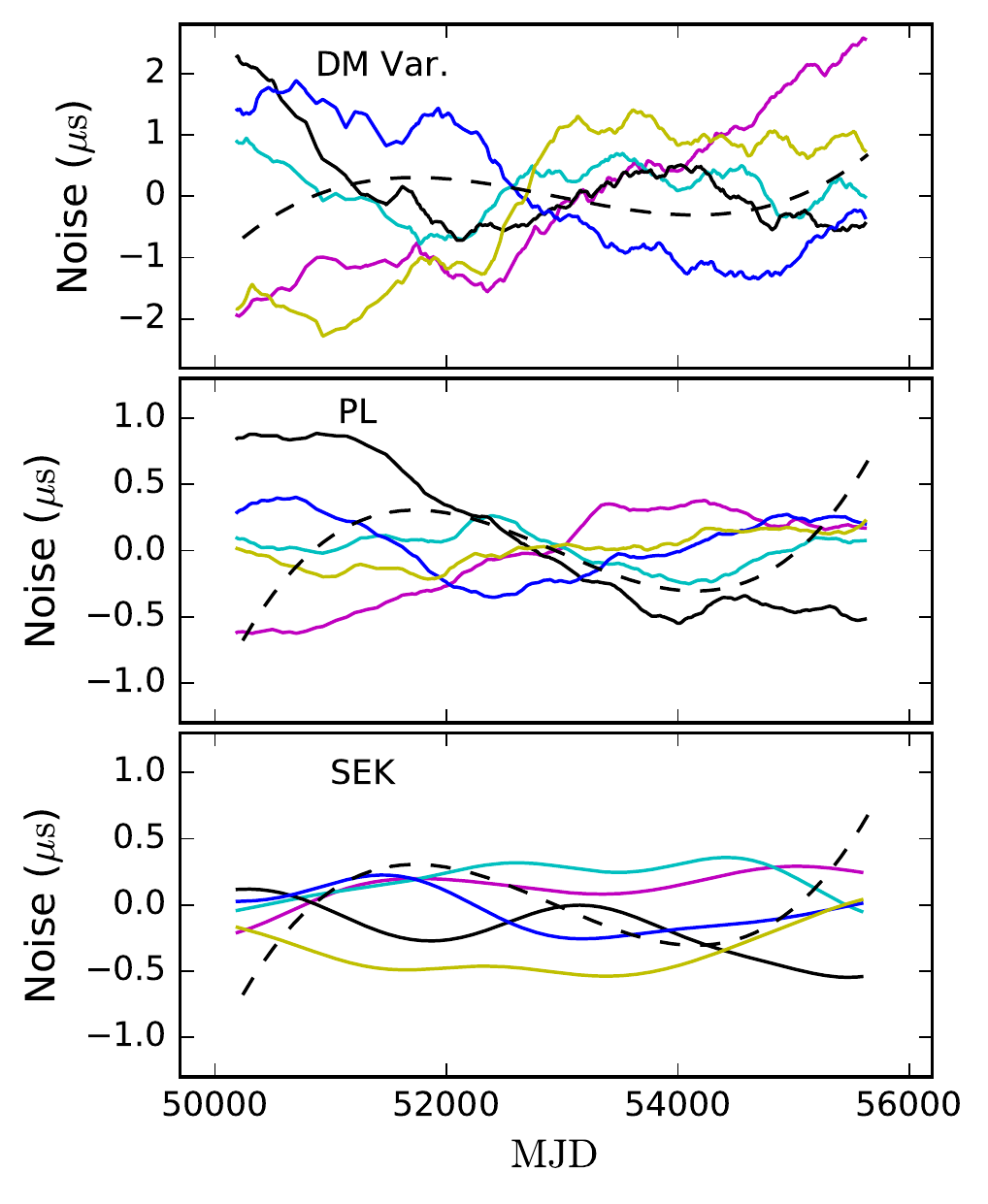}
 \caption{Representative examples of the simulations of correlated timing noise. The top panel shows the contribution to observations at 1400 MHz from DM variations, while the middle and bottom panels show contributions from the power law and squared exponential kernel noise, respectively. Each panel contains five realisations which are represented by lines of different colours. The black, dashed line in each panel shows expected timing signal of $\ddot{\nu}_{\rm in}$.} 
 \label{fig:dm_sig}
\end{figure}

We added $\ddot{\nu}_{\rm in}$ to the mock TOAs by adding in the pulsar ephemeris a value of $\ddot{\nu}_{\rm in}=1.4\times10^{-28}$\,s$^{-3}$, which is larger than the typical value for this pulsar due to the radial velocity \citep{lbs18} and makes the fitting results statistically more significant. The black, dashed line in Fig.~\ref{fig:dm_sig} shows the expected timing signal of $\ddot{\nu}_{\rm in}$.

\subsection{The DM variations}
\label{DMVar}
Assuming Kolmogorov turbulence in the interstellar plasma, the power spectrum of the DM variations can be modeled by a power law \citep{kcs+13}
\begin{equation}
 \label{DM_PL}
P(f)=0.0112 D(\tau) \tau^{-5/3}f^{-8/3},
\end{equation}
where $P(f)$ is the power density at the DM fluctuation frequency $f$, and in unit of yr$^3$, $\tau$ is the time lag and $D(\tau)$ is the structure function, which is the autocorrelation of time delays caused by DM. We used a value of $D(\tau)=1.6$\,$\mu$s$^2$ for a lag of $\tau=1000$ days taken from PSR J0437$-$4715 \citep[table 2]{kcs+13}. 

To generate the time series of DM variations, we Fourier transformed a set of complex Gaussian numbers with both the real and imaginary parts having zero mean and unit variance. The modulus of the complex numbers was then scaled by $\sqrt{P(f)}$ to reflect the amplitude of the DM variations. To avoid the loss of power in the low frequency part of the spectrum during the Fourier transformation, we generated time series that are at least 100 times longer than the desired length. This long time series can then be cut into 15-year segments which are used as a separate realisation of the DM variations. All of these treatments have been well integrated into the \textsc{tempo2} plug-in \texttt{addDmVar}, which was used to inject the signals of DM variation into the mock TOAs. The top panel of Fig.~\ref{fig:dm_sig} shows five representative examples of DM variations generated by our simulation.

\subsection{The red noise model}
The power law parameterization of red noise is ubiquitous among the analysis of timing noise (e.g. \citealp{cll+16,rhc+16,abb2+18}, and this work). However, the underlying process behind the red noise model is unknown and may not be best described by a power law. If the red noise model used in the analysis does not match the underlying one, parameter and error estimation may become inaccurate. In order to assess the impact of this on our work, we simulate two commonly used red-noise models to produce two different sets of mock TOAs, and analyze both with the same method.

Firstly, we used the model described by \citet{chc+11} which parameterizes the power spectrum of the red noise by a power law given by
\begin{equation}
 \label{RedPL}
 P(f)=P_0\bigg[1+\bigg(\frac{f}{f_{\rm c}}\bigg)^2\bigg]^{-\alpha/2},
\end{equation}
where $P_0$ is the amplitude of power, $f_{\rm c}$ the corner frequency and $\alpha$ the power law index. In the simulations we set $P_0=10^{-25}$\,yr$^3$, $f_{\rm c}=0.01$\,yr$^{-1}$ and $\alpha=3$, which are typical values for the red noise models of pulsars in the PPTA data set \citep[table 2]{rhc+16}. The middle panel of Fig.~\ref{fig:dm_sig} shows five representative examples of power law red noise generated by our simulation. The power law red noise was added into the mock TOAs by using the \textsc{tempo2} plug-in \texttt{addRedNoise}.

For the second noise model, we generate red noise in the time domain using the squared exponential kernel commonly used for Gaussian process regression (e.g. \citealp{rw05}). This states that the covariance between two data points at times $t_1$ and $t_2$ is given by
\begin{equation}
 \label{RBFcovar}
 {\rm cov}(t_1, t_2)=A\exp\bigg[-\frac{(t_1-t_2)^2}{2l^2}\bigg],
\end{equation}
where $A$ is a constant describing the strength of the correlation and $l$ is the time-scale of correlation. We used $A=0.05\,\mu$s$^2$ and $l=1000$ days to give a similar magnitude of red noise to that in the power law model. To simulate this noise, a covariance matrix, \textbf{C}, was computed using Eqn.~\ref{RBFcovar}, and was then decomposed to a lower triangular matrix \textbf{L} by the Cholesky decomposition $\mathbf{C}=\mathbf{L}\mathbf{L}^\intercal$. Finally the red noise is given by $\mathbf{L}w$, where $w$ is a vector of Gaussian white noise with unit variance. This is the inverse of the process used to whiten the correlated data described by \cite{chc+11}. The bottom panel of Fig.~\ref{fig:dm_sig} shows five representative examples of squared exponential kernel noise generated by our simulation.

\begin{figure}
 \includegraphics[width=8.5cm]{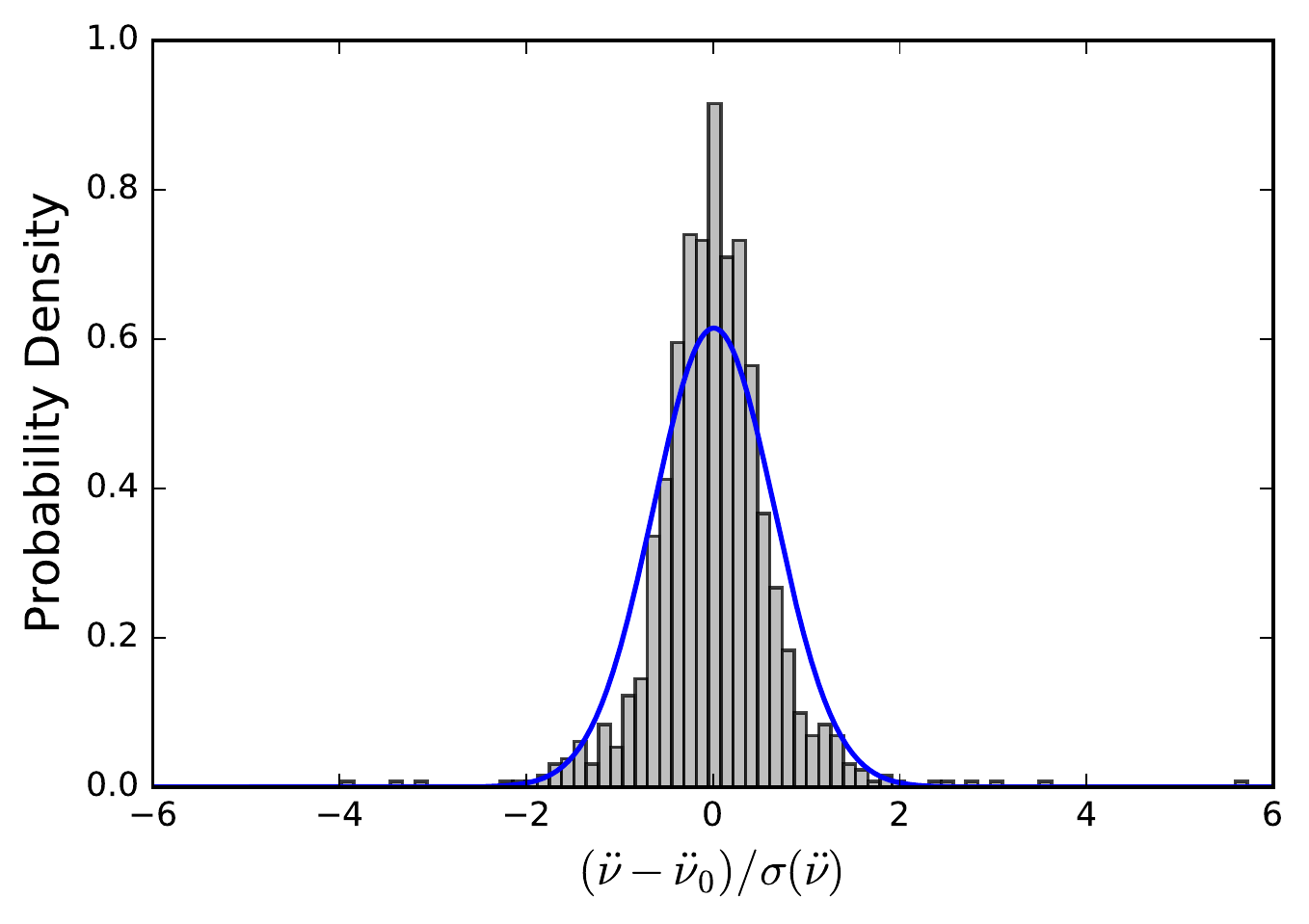}
  \caption{The fitting results for mock TOAs containing power law red noise: The histogram of the normalized $\ddot{\nu}$ and the normal distribution (blue line) used to fit the histogram are shown. For the fit, the mean is $\mu=0.01$ and standard deviation is $\sigma=0.65$.}
  \label{FitCheck}
\end{figure}

\subsection{Validity of the method}
We normalized the mean of $\ddot\nu$ of each realisation by the corresponding uncertainty, $\sigma(\ddot\nu)$, through $(\ddot\nu-\ddot\nu_{\rm in})/\sigma(\ddot\nu)$. A histogram of the normalized quantity was then plotted and fitted with a normal distribution. 

For the power law case, as can be seen from Fig.\,\ref{FitCheck}, the normalized $\ddot{\nu}$ can approximately be fitted with a normal distribution with zero mean and a small $\sigma$ of 0.65. The general consistency between the histogram and the normal distribution proves the validity of our analysis method, although the standard deviation differs from $\sigma=1$ of the ideal case which suggests a slight overestimation of the measurement errors. 

For the second case, where the data are simulated using the squared exponential kernel, but fit assuming a power law process, the results are not so well fit by a normal distribution due to a small number of outliers
($\sim 1$ per cent of the simulations) with $(\ddot\nu-\ddot\nu_{\rm in})/\sigma(\ddot\nu)>5$. Using the power law model to fit for squared exponential kernel noise is not guaranteed to give correct answers, but as the small number of outliers show, this method can recover the injected $\ddot{\nu}$ very well. The outliers may be a compound effect of the mismatch between the noise models and the very smooth noise generated by the squared exponential kernel. We also find that for these outliers, the red noise generated by the squared exponential kernel is very similar to the signal of $\ddot{\nu}$, with little power at higher frequencies, and so not well modelled by the power law noise model. This leads the algorithm to attribute most of the noise power to $\ddot{\nu}$ and significantly underestimate the associated error. When these outliers are removed, the measurements are well fit by a normal distribution with mean of $-$0.01 and standard deviation of 0.67, consistent with the results from the power law red noise.

For each realisation of the two red-noise models, we also computed the confidence interval of 95 per cent confidence level (C.L.). We then counted the number of intervals that contain $\ddot\nu_{\rm in}$ and calculated the ratio of this number to the total number of intervals. In the simulation of power law noise, the ratio is 98.7 per cent, while in the simulation of squared exponential kernel noise, the ratio is 96.7 per cent\footnote{This number is not affected by the inclusion or exclusion of the  outliers.}. Both ratios are a little higher than the ideal value of 95 per cent. 

In summary, we find that our method returns reasonable estimates of $\ddot{\nu}$ and its error, though the error may be over estimated by up to a factor of 2. We also find that the method is robust against a mismatch between the red noise models (in this case, using a power law noise model to fit for squared exponential noise), although we observed a small number of outliers ($\sim$ 1 per cent).
This analysis does not exhaust all possible models for pulsar red noise, however, it gives us confidence in the robustness of our measurement and error estimates.

\section{Timing data}
\label{data}
We applied the method described in Section\,\ref{fitting} to the timing data from the EPTA \citep{dcl+16} and PPTA data releases \citep{rhc+16} separately. We did not include the data set from NANOGrav \citep{abb2+18}, as this data set has a comparable time span but much larger number of TOAs, which can increase the sensitivity to $\ddot{\nu}$ but would make the fitting process computationally more expensive and require a different treatment. Our analyses thus included 49 MSPs, among which 13 MSPs have timing data from both the EPTA and PPTA (as seen in Table~\ref{F2ms}). 

We also included additional timing data of two pulsars, PSRs B1855+09 and B1937+21, consisting of observations recorded by the Arecibo radio telescope between $\sim1986$ and 1993 \citep{ktr94}. For both pulsars, in addition to the analysis of the EPTA and PPTA data sets, a joint analysis was carried out by combining the data from \cite{ktr94} with those from the EPTA.

\begin{figure*}
 \includegraphics[trim=15 44 15 60, clip, width=18.5cm]{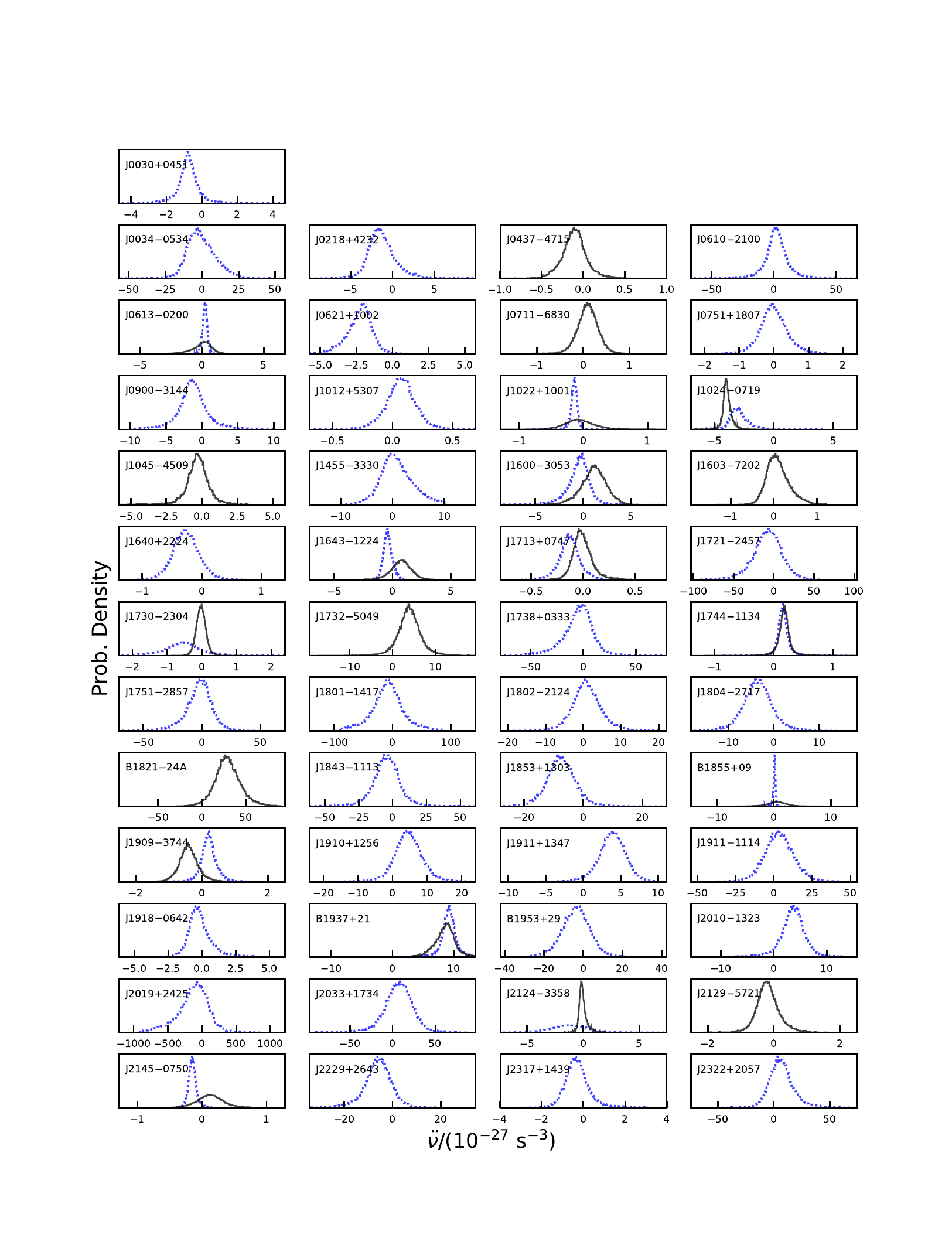}
  \caption{The probability density function of $\ddot{\nu}$ of the 49 pulsars listed in the EPTA and PPTA data release. The horizontal axes are for $\ddot{\nu}$ in units of $10^{-27}$\,s$^{-3}$, while the vertical axes are for the normalised probability density. The blue, dotted curves are produced from the EPTA data release \citep{dcl+16}, while the black, solid ones from the PPTA \citep{rhc+16}. Pulsars with data from both the EPTA and PPTA have two PDFs.} 
  \label{F2pdf}
\end{figure*}

\begin{table*}
    \centering
    \scriptsize
    \caption{Characterisations of $\ddot{\nu}$, red noise and timing baseline of the EPTA and PPTA MSPs: The confidence interval with 95 per cent C.L., mean, standard deviation (assuming a normal distribution), significance of $\ddot{\nu}$, predicted $\ddot{v}$ due to radial velocity ($\ddot{\nu}_\parallel$), power index of red noise ($\alpha$) used in the computations, current timing baseline ($T_0$). The predicted total timing baselines to measure $\ddot{\nu}_\parallel$ for the MSPs in the EPTA and PPTA are labeled by $T_1$ (assuming white noise) and $T_2$ (using the listed red noise parameters in the column of $\alpha$). Symbols E and P in the last column are for EPTA and PPTA respectively. The pulsars observed by both PTAs have two sets of values, each for one PTA. $\ddot{\nu}_\parallel$ with $\dagger$ was computed from an observed radial velocity \protect\citep[table 2]{lbs18}, otherwise we assume a value of 50\,km\,s$^{-1}$. $\alpha$ marked with $\star$ was from \protect\citet[table 2]{cll+16} or \protect\citet[table 2]{rhc+16}, otherwise we assume a value of 3.}

     \begin{tabular}{lcrrrrlrccc}
      \hline
       \hline
 \multicolumn{1}{l}{PSR} & \multicolumn{1}{c}{95\% C.L.} & \multicolumn{1}{c}{$\ddot{\nu}$}  & \multicolumn{1}{c}{$\sigma(\ddot{\nu})$} & \multicolumn{1}{c}{Sig.} & \multicolumn{1}{c}{$\ddot{\nu}_\parallel$} & \multicolumn{1}{c}{$\alpha$} & \multicolumn{1}{c}{$T_0$} & \multicolumn{1}{c}{$T_1$} & \multicolumn{1}{c}{$T_2$} & \multicolumn{1}{c}{PTA}  \\
    \cline{2-4}  \cline{9-10}  
     \\[-0.8em]
    & \multicolumn{3}{c}{$10^{-27}$\,s$^{-3}$} &  & \multicolumn{1}{r}{$10^{-30}$\,s$^{-3}$} & & \multicolumn{1}{c}{yr} & \multicolumn{2}{c}{$10^2$ yr} & \\  
       \hline

  J0030$+$0451  &  $(  -2.9 $, $   0.9 )$ & $   -0.8 $ &    0.9 &     0.9 &   $ -0.08           $ & $ 5.2^\star  $ &   15.1  &  $ -   $   & $ 750 $ &  E \\ 
  J0034$-$0534  &  $(  -20  $, $   23  )$ & $    0   $ &     11 &       0 &   $ -0.93           $ & $ 3.0   	  $ &   13.5  &  $ 3.1 $   & $ 33  $ &  E \\  
  J0218$+$4232  &  $(  -4.5 $, $   3.4 )$ & $   -1   $ &      2 &     0.6 &   $ -0.19           $ & $ 3.9^\star  $ &   17.6  &  $ -   $   & $ 120 $ &  E \\  
  J0437$-$4715  &  $(  -0.4 $, $   0.2 )$ & $   -0.1 $ &    0.2 &     0.7 &   $ -41             $ & $ 3.0^\star  $ &   14.9  &  $ -   $   & $ 46  $ &  P \\  
  J0610$-$2100  &  $(  -29  $, $   27  )$ & $    1   $ &    13  &     0.1 &   $ -1.1            $ & $ 2.7^\star  $ &   6.9   &  $ -   $   & $ 12  $ &  E \\  
  J0613$-$0200  &  $(  -0.2 $, $   0.7 )$ & $    0.2 $ &    0.2 &     1.1 &   $ -0.43           $ & $ 4.1^\star  $ &   16.1  &  $ -   $   & $ 22  $ &  E \\  
                &  $(  -4.8 $, $   2.5 )$ & $    -0.2$ &    1.7 &     0.1 &   $                 $ & $ 5.0^\star  $ &   11.2  &  $ -   $   & $ 270 $ &  P \\  
  J0621$+$1002  &  $(  -5.1 $, $   -0.6)$ & $   -2.4 $ &    1.1 &     2.1 &   $ -0.004          $ & $ 2.4^\star  $ &   11.8  &  $ -   $   & $ 59  $ &  E \\  
  J0711$-$6830  &  $(  -0.5 $, $   0.6 )$ & $    0.1 $ &    0.3 &     0.3 &   $ -0.96           $ & $ 3.0 	     $ &   17.1  &  $ 1.4 $   & $ 390 $ &  P \\  
  J0751$+$1807  &  $(  -0.9 $, $   1.0 )$ & $    0   $ &    0.5 &       0 &   $ -0.63           $ & $ 3.0^\star  $ &   17.6  &  $ -   $   & $ 11  $ &  E \\  
  J0900$-$3144  &  $(  -6.4 $, $   3.7 )$ & $   -1.3 $ &    2.3 &     0.6 &   $ -0.01           $ & $ 3.0 	     $ &   6.9   &  $ 4.4 $   & $ 99  $ &  E \\  
  J1012$+$5307  &  $(  -0.21$, $   0.34)$ & $    0.07$ &    0.14&     0.5 &   $  1.3^\dagger    $ & $ 1.7^\star  $ &   16.8  &  $ -   $   & $ 1.9 $ &  E \\  
  J1022$+$1001  &  $(  -0.26$, $  -0.04)$ & $   -0.14$ &    0.06&     2.4 &   $ -0.24           $ & $ 1.6^\star  $ &   17.5  &  $ -   $   & $ 2.6 $ &  E \\  
                &  $(  -0.5 $, $   0.8 )$ & $    0   $ &    0.3 &     0.1 &   $                 $ & $ 3.0 	     $ &   8.13  &  $ 1.0 $   & $ 14  $ &  P \\  
  J1024$-$0719  &  $(  -4.3 $, $  -0.7 )$ & $   -2.9 $ &    0.9 &     3.1 &   $ 30^\dagger      $ & $ 3.9^\star  $ &   17.3  &  $ -   $   & $ -   $ &  E \\  
                &  $(  -4.7 $, $  -2.9 )$ & $   -3.9 $ &    0.5 &     8.5 &   $                 $ & $ 6.0^\star  $ &   15.1  &  $ -   $   & $ -   $ &  P \\  
  J1045$-$4509  &  $(  -2.7 $, $   1.5 )$ & $   -0.3 $ &    1.0 &     0.4 &   $ -0.10           $ & $ 3.0^\star  $ &   17.0  &  $ -   $   & $ 2200$ &  P \\  
  J1455$-$3330  &  $(  -5.3 $, $   7.8 )$ & $    0.8 $ &    3.2 &     0.3 &   $ -0.10           $ & $ 3.6^\star  $ &   9.2   &  $ -   $   & $ 80  $ &  E \\  
  J1600$-$3053  &  $(  -4.5 $, $   1.8 )$ & $   -0.6 $ &    1.5 &     0.4 &   $ -0.16           $ & $ 1.7^\star  $ &   7.6   &  $ -   $   & $ 4.8 $ &  E \\  
   	          &  $(  -2.1 $, $   3.7 )$ & $    1.1 $ &    1.4 &     0.7 &   $                 $ & $ 2.5^\star  $ &   9.1   &  $ -   $   & $ 9.5 $ &  P \\  
  J1603$-$7202  &  $(  -0.4 $, $   0.8 )$ & $    0.1 $ &    0.3 &     0.3 &   $ -0.05           $ & $ 2.5^\star  $ &   15.3  &  $ -   $   & $ 1100$ &  P \\  
  J1640$+$2224  &  $(  -0.8 $, $   0.3 )$ & $   -0.3 $ &    0.3 &     0.9 &   $ -0.49           $ & $ 0.4^\star  $ &   17.3  &  $ -   $   & $ 2.0 $ &  E \\  
  J1643$-$1224  &  $(  -1.2 $, $   0.6 )$ & $   -0.4 $ &    0.4 &     0.9 &   $ -0.14           $ & $ 1.7^\star  $ &   17.3  &  $ -   $   & $ 7.2 $ &  E \\  
  	             &  $(  -2.0 $, $   3.0 )$ & $    0.7 $ &    1.2 &     0.6 &   $                 $ & $ 4.0^\star  $ &   17.0  &  $ -   $   & $ 120 $ &  P \\  
  J1713$+$0747  &  $(  -0.4 $, $   0.1 )$ & $   -0.1 $ &    0.1 &     0.9 &   $ -0.10           $ & $ 5.4^\star  $ &   17.7  &  $ -   $   & $ 230 $ &  E \\  
  	             &  $(  -0.2 $, $   0.3 )$ & $    0   $ &    0.1 &       0 &   $                 $ & $ 2.0^\star  $ &   17.0  &  $ -   $   & $ 6.3 $ &  P \\  
  J1721$-$2457  &  $(  -53  $, $    32 )$ & $   -8   $ &    21  &     0.4 &   $ -2.1            $ & $ 1.9^\star  $ &   12.7  &  $ -   $   & $ 9.5 $ &  E \\  
  J1730$-$2304  &  $(  -2.5 $, $   0.9 )$ & $   -0.6 $ &    0.9 &     0.7 &   $ -0.74           $ & $ 2.9^\star  $ &   16.7  &  $ -   $   & $ 12  $ &  E \\  
  	             &  $(  -0.3 $, $   0.3 )$ & $    0   $ &    0.2 &     0.1 &   $                 $ & $ 3.0 	     $ &   16.9  &  $ 1.2 $   & $ 5.5 $ &  P \\  
  J1732$-$5049  &  $(  -2.7 $, $  10.4 )$ & $    3.9 $ &    3.2 &     1.2 &   $ -0.22           $ & $ 2.0^\star  $ &   8.0   &  $ -   $   & $ 970 $ &  P \\  
  J1738$+$0333  &  $(  -42  $, $    22 )$ & $   -5   $ &    16  &     0.3 &   $ -0.13^\dagger   $ & $ 3.0 	     $ &   7.3   &  $ 3.3 $   & $ 56  $ &  E \\  
  J1744$-$1134  &  $(   0   $, $   0.5 )$ & $    0.2 $ &     0.2&     1.1 &   $ -1.3            $ & $ 2.7^\star  $ &   17.3  &  $ -   $   & $ 3.4 $ &  E \\  
  	             &  $(  -0.1 $, $   0.6 )$ & $    0.2 $ &    0.2 &     0.8 &   $                 $ & $ 3.0 	     $ &   16.1  &  $ 1.1 $   & $ 5.2 $ &  P \\  
  J1751$-$2857  &  $(  -37  $, $   21  )$ & $   -3   $ &   14   &     0.2 &   $ -0.22           $ & $ 3.0 	     $ &   8.3   &  $ 3.1 $   & $ 46  $ &  E \\  
  J1801$-$1417  &  $( -69   $, $    55 )$ & $   -8   $ &   28   &     0.3 &   $ -0.42           $ & $ 3.3^\star  $ &   7.1   &  $ -   $   & $ 61  $ &  E \\  
  J1802$-$2124  &  $(  -7.7 $, $  10.6 )$ & $    0.9 $ &    4.4 &     0.2 &   $ -0.01           $ & $ 2.3^\star  $ &   7.2   &  $ -   $   & $ 41  $ &  E \\  
  J1804$-$2717  &  $(  -11  $, $     4 )$ & $    -4  $ &     4  &     1.0 &   $ -0.37           $ & $ 3.0 	     $ &   8.1   &  $ 1.8 $   & $ 18  $ &  E \\  
  B1821$-$24A   &  $(  -1   $, $    63 )$ & $   29   $ &   16   &     1.9 &   $ -               $ & $ 3.5^\star  $ &   5.8   &  $ -   $   & $ -   $ &  P \\  
  J1843$-$1113  &  $( -28   $, $   20  )$ & $   -4   $ &   12   &     0.3 &   $ -0.09           $ & $ 1.5^\star  $ &   10.1  &  $ -   $   & $ 15  $ &  E \\  
  J1853$+$1303  &  $( -18   $, $    3  )$ & $   -8   $ &   6    &     1.4 &   $ -0.03           $ & $ 3.0 	     $ &   8.4   &  $ 4.3 $   & $ 80  $ &  E \\  
  B1855$+$09    &  $(  -0.6 $, $   0.5 )$ & $    0.1 $ &    0.3 &     0.3 &   $ -0.08           $ & $ 2.4^\star  $ &   17.3  &  $ -   $   & $ 13  $ &  E \\  
  	             &  $(  -4.5 $, $   6.1 )$ & $    0.6 $ &    2.4 &     0.2 &   $                 $ & $ 3.0        $ &   6.9   &  $ 2.1 $   & $ 67  $ &  P \\  
  J1909$-$3744  &  $(  -0.4 $, $   0.8 )$ & $    0.2 $ &    0.3 &     0.7 &   $ -8.0^\dagger    $ & $ 2.3^\star  $ &   9.4   &  $ -   $   & $ 0.9 $ &  E \\  
  	             &  $(  -1.0 $, $   0.4 )$ & $    -0.4$ &    0.4 &     1.0 &   $                 $ & $ 2.0^\star  $ &   8.2   &  $ -   $   & $ 0.9 $ &  P \\  
  J1910$+$1256  &  $(  -5   $, $   14  )$ & $    5   $ &    5   &     1.0 &   $ -0.13           $ & $ 3.0  	     $ &   8.5   &  $ 2.7 $   & $ 36  $ &  E \\  
  J1911$+$1347  &  $(  -0.6 $, $   7.8 )$ & $     3.9$ &    2.2 &     1.7 &   $ -0.06           $ & $ 3.0  	     $ &   7.5   &  $ 2.4 $   & $ 33  $ &  E \\  
  J1911$-$1114  &  $(  -16  $, $    25 )$ & $    4   $ &    11  &     0.3 &   $ -0.89           $ & $ 3.0  	     $ &   8.8   &  $ 2.1 $   & $ 22  $ &  E \\  
  J1918$-$0642  &  $(  -1.7 $, $   3.1 )$ & $    0   $ &    1   &       0 &   $ -0.13           $ & $ 5.4^\star  $ &   12.8  &  $ -   $   & $ 770 $ &  E \\  
  B1937$+$21    &  $(   4.6 $, $  12.5 )$ & $    9   $ &    2   &     4.8 &   $ -0.001          $ & $ 6.2^\star  $ &   24.1  &  $ -   $   & $ -   $ &  E \\  
  	             &  $(   4.2 $, $  12.9 )$ & $    8   $ &    2   &     3.8 &   $                 $ & $ 4.5^\star  $ &   15.5  &  $ -   $   & $ -   $ &  P \\  
  B1953$+$29    &  $(   -20 $, $    12 )$ & $   -4   $ &    8   &     0.4 &   $ -0.04           $ & $ 3.0 	     $ &   8.1   &  $ 4.2 $   & $ 79  $ &  E \\  
  J2010$-$1323  &  $(  -3.8 $, $   9.3 )$ & $    3   $ &    3   &     1.1 &   $ -0.09           $ & $ 3.0  	     $ &   7.4   &  $ 2.3 $   & $ 31  $ &  E \\  
  J2019$+$2425  &  $(  -688 $, $   323 )$ & $   -123 $ &    240 &     0.5 &   $ -1.4            $ & $ 3.0  	     $ &   9.1   &  $ 4.5 $   & $ 83  $ &  E \\  
  J2033$+$1734  &  $(   -39 $, $    42 )$ & $    6   $ &    19  &     0.3 &   $ -0.23           $ & $ 3.0  	     $ &   7.9   &  $ 3.2 $   & $ 50  $ &  E \\  
  J2124$-$3358  &  $(  -5.7 $, $   3.2 )$ & $   -1   $ &     2  &     0.6 &   $ -6.5            $ & $ 3.0  	     $ &   9.4   &  $ 0.8 $   & $ 3.8 $ &  E \\   
  	             &  $(  -0.7 $, $   1.2 )$ & $    0   $ &    0.5 &       0 &   $                 $ & $ 3.5^\star  $ &   16.8  &  $ -   $   & $ 2.5 $ &  P \\  
  J2129$-$5721  &  $(  -0.9 $, $   0.7 )$ & $   -0.2 $ &    0.4 &     0.4 &   $ -0.56           $ & $ 1.0^\star  $ &   15.4  &  $ -   $   & $ 160 $ &  P \\  
  J2145$-$0750  &  $(  -0.28$, $  0.03 )$ & $   -0.14$ &   0.08 &     1.8 &   $ -0.13           $ & $ 4.1^\star  $ &   17.5  &  $ -   $   & $ 27  $ &  E \\  
  	             &  $(  -0.4 $, $   0.8 )$ & $    0.2 $ &   0.3  &     0.5 &   $                 $ & $ 4.0^\star  $ &   16.7  &  $ -   $   & $ 58  $ &  P \\  
  J2229$+$2643  &  $( -21.8 $, $   6.5 )$ & $   -6   $ &    7   &     0.9 &   $ -0.15           $ & $ 3.0 	 	  $ &   8.2   &  $ 2.8 $   & $ 39  $ &  E \\  
  J2317$+$1439  &  $(  -1.5 $, $   1.7 )$ & $   -0.2 $ &   0.8  &     0.3 &   $ -0.04           $ & $ 3.0 	     $ &   17.3  &  $ 4.5 $   & $ 52  $ &  E \\  
  J2322$+$2057  &  $(  -18  $, $    42 )$ & $   7    $ &   15   &     0.5 &   $ -1.4            $ & $ 3.0 	     $ &   7.9   &  $ 1.8 $   & $ 18  $ &  E \\  

  \hline
  \end{tabular}
  \label{F2ms}
\end{table*}

\section{Results and Discussions}
\label{results}

\subsection{$\ddot{\nu}$ in real timing data}
After marginalizing all the other variables like white noise, red noise and DM variations, we show in Fig.~\ref{F2pdf} the normalized probability density functions (PDFs) of $\ddot{\nu}$ for the pulsars observed by the EPTA and PPTA. In Fig.~\ref{F2pdf}, the PDFs shown by the blue, dotted line are obtained by using the EPTA data, while those shown by the black, solid line are from the PPTA data. For the pulsars with data from both PTAs, the PDFs are generally consistent, although they may be different in peak position and/or width, which is directly affected by the timing precision, cadence, time span and noise level of the pulsar.

One can see that the PDFs  of most MSPs are approximately symmetric and normal. The most probable value of $\ddot{\nu}$, which is the value that corresponds to the peak of a PDF, is thus generally consistent with the mean value. Using Fig.~\ref{F2pdf}, we find that the most probable value of $\ddot{\nu}$ of only two MSPs, PSRs~J1024$-$0719 and B1937+21, deviate from zero significantly, although PSRs~J0621+1002 and J1022+1001 also seem to have non-zero $\ddot{\nu}$. 

We compute the standard deviation of each PDF and treat this as an estimate of the measurement uncertainty, $\sigma(\ddot{\nu})$, see Table~\ref{F2ms}. This treatment thus assumes that all the PDFs are normal distributions. Of all the 49 MSPs, PSR~J1022+1001 has the smallest measurement uncertainty, which is $\sigma(\ddot{\nu})= 6\times10^{-29}$\,s$^{-3}$. 
Except for the case of PSR~J0437$-$4715, this value is still a few times larger than the largest possible $\ddot{\nu}$ which could be induced by the radial velocity of this sample of pulsars \citep{lbs18}.

The values of $\sigma(\ddot{\nu})$ in Table~\ref{F2ms} are generally larger than those predicted in \cite{lbs18}, where white noise, even cadence and constant timing precision were assumed. We think that the increase of $\sigma(\ddot{\nu})$ is mainly caused by the inclusion of red noise and DM variations, although the uneven cadence and varying precision of each TOA also contribute. We thus need to check previous significant detections of $\ddot{\nu}$ in the references (see Table~\ref{SigF2}), which did not account for either DM variations or red noise. 

To quantify the detection significance of $\ddot{\nu}$, it is convenient to use as an indicator the significance, which is defined here as the ratio of the mean to $\sigma(\ddot{\nu})$. In Table~\ref{F2ms}, the significance of all but two MSPs (PSRs~J0621+1002, J1022+1001, J1024$-$0719 and B1937+21) are smaller than 2, and we consider these statistically insignificant. It is therefore more meaningful to use the confidence interval to constrain $\ddot{\nu}$. The confidence intervals with 95 per cent C.L. are given in Table~\ref{F2ms}.

Let us further consider our statistically significant measurements. We will also consider previously published significant measurements of $\ddot{\nu}$ (PSRs~J1012+5307, J1024$-$0719, B1821$-$24A, B1855+09 and B1937+21). According to \citet[section 3.3.3]{van03a} and \citet[section 4]{lbs18}, PSRs~J0437$-$4715 may also have a high and measurable $\ddot{\nu}$. We present the properties of the PDFs of these pulsars in Table~\ref{SigF2}. PSRs~J0613$-$0200 and J1909$-$3744 are also discussed due to their unusual characteristics. 

\begin{table}
 \centering
  \caption{The MSPs with potentially significant $\ddot{\nu}$ and their statistical quantities. Columns are confidence interval with 95 per cent C.L., mean value, standard deviation (assuming a normal PDF) and significance (only value much larger than 1 is listed). Symbols E and P in the last column refer to the EPTA and PPTA data set respectively, and EK represents the data combining those from the EPTA and \protect\cite{ktr94}. Lines labelled with E, P or EK are new results obtained by this paper. The remaining references are: (1) \protect\cite{lcw+01}, (2) \protect\cite{bjs+16}, (3) \protect\cite{kkn+16}, (4) \protect\cite{cbl+96}, (5) \protect\cite{jgk+13}, (6) \protect\cite{ktr94}.}
  \begin{tabular}{lrrccc}
  \hline\hline
      \multicolumn{1}{l}{PSR} & \multicolumn{1}{c}{95\% C.L.} & \multicolumn{1}{c}{$\ddot{\nu}$}  & \multicolumn{1}{c}{$\sigma(\ddot{\nu})$} & \multicolumn{1}{c}{Sig.} & \multicolumn{1}{c}{ref.} \\
      \cline{2-4}
      \\[-0.8em]
    & \multicolumn{3}{c}{$10^{-27}$\,s$^{-3}$} &   & \\
    \hline

  J0437$-$4715 & $(  -0.43 $, $   0.24 )$ & $     -0.11 $ &     0.17 &     $-$  &      P \\
  \\
  J1012+5307 & $(  -0.21 $, $   0.34 )$ & $      0.07 $ &      0.14 &    $-$   &      E \\
             & & $-9.8$ & 2.1 & 4.7 & (1) \\
 \\
 J1024$-$0719 & $(  -4.28 $, $  -0.69 )$ & $     -2.87 $ &      0.93 &     3.1 &      E \\
 & $(  -4.74 $, $  -2.94 )$ & $                    -3.93 $  &      0.46 &     8.5 &      P \\
 & & $-3.92$ & 0.02 & 196 & (2) \\
 & & $-4.1$ & 1.0 & 4.1 & (3) \\
 \\
B1821$-$24A & $(  -1.24 $, $  63.17 )$ & $     29.42 $ &    15.75 &     1.9 &      P \\
 & & $-175$ & 7 & 25.0 & (4) \\
 & & $-26.49$ & 0.05 & 529.8 & (5) \\

  \\
  B1855+09 & $(  -0.56 $, $   0.46 )$ & $      0.08 $ &      0.27 &  $-$    &      E \\
 & $(  -4.49 $, $   6.13 )$ & $                     0.61 $  &      2.43 &  $-$    &      P \\
 & $( -0.26$, $0.04) $ & $-0.11$ & 0.08 & $-$ & EK \\
 & & $-1.0$ & 0.9 & $-$ & (6) \\
 \\
B1937+21 & $(   4.61 $, $  12.48 )$ & $      9.08 $ &      1.89 &     4.8 &      E \\
 & $(   4.16 $, $  12.90 )$ & $                     8.49 $  &      2.25 &     3.8 &      P \\  
 & $(4.40$, $14.82)$ & 10.41 & 2.60 & 4.0 & EK \\
 & & 13.2 & 0.3 & 44.0 & (6) \\
 
  \hline
  \end{tabular}
  \label{SigF2}
\end{table}

\subsubsection{PSR\,J0437$-$4715}
\label{J0437}
In the absence of red noise, PSR\,J0437$-$4715 would have a detectable $\ddot{\nu}$ if its radial velocity exceeds 33\,km\,s$^{-1}$ \citep{lbs18}. We find that the measurement uncertainty of $\ddot{\nu}$ increases by a factor 40 when contributions from DM variations and timing noise are taken into account. Hence $\ddot{\nu}$ is undetectable at present. The current constraint on $\ddot{\nu}$ is ($-4.3$, 2.4)$\times10^{-28}$\,s$^{-3}$ (95 per cent C.L.), it thus constrains the radial velocity to ($-527$, 294)\,km\,s$^{-1}$.

\subsubsection{PSR\,J0613$-$0200}
A glitch of 0.82 nHz in the spin frequency and $-1.6\times10^{-19}$\,Hz~s$^{-1}$ in the spin frequency derivative at MJD  $50888\pm30$ has been reported by \cite{mjs+16} for PSR~J0613$-$0200, but the earliest EPTA data in this analysis were taken a few days after the glitch and the earliest PPTA data $\sim600$ days after. \citeauthor{mjs+16} do not find any evidence of a glitch recovery in this pulsar, as may be expected given its age \citep{lps95}. We also see no evidence for a glitch recovery, thus our constraint on $\ddot{\nu}$ can be directly used to constrain the magnitude of $\ddot{\nu}$ of other origins.

\subsubsection{PSR\,J0621+1002}
This pulsar has a $\ddot{\nu}$ in the range of $(-5.1, -0.6)\times10^{-27}{\rm s}^{-3}$ (95 per cent C.L.), although contributions from the pulsar radial velocity, the radiation braking and the Galactic potential \citep{lbs18} cannot explain a $\ddot{\nu}$ of such large magnitude. Since the power spectrum of the red noise for this object is relatively shallow ($\alpha$=2.4, \citealt{dcl+16}) and the current timing baseline is relatively short ($\sim 12$ years), $\sigma(\ddot{\nu})$ will further decrease by a noticeable amount when the timing baseline increases (see Section~\ref{scaling_expression}) at which we will probably be able to reject or accept the detection.

\subsubsection{PSR\,J1012+5307}
For the $\ddot{\nu}$ of PSR\,J1012+5307, \cite{lcw+01} reported a value of $(-9.8\pm2.1)\times10^{-27}$\,s$^{-3}$, but \cite{lwj+09} did not measure the $\ddot{\nu}$. Our analysis gives a tight constraint of $(-2.1, 3.4)\times10^{-28}$\,s$^{-3}$ with 95 per cent C.L.. We thus can not confirm the measurement. Note that the power index of red noise in this pulsar is very small (Table~\ref{F2ms}), but it is still difficult to measure the predicted $\ddot{\nu}=(1.3\pm0.2)\times10^{-30}$\,s$^{-3}$ \citep{lcw+01, lbs18} in the near future. The resultant constraint on the radial velocity will still be loose, although a measurement of $(44\pm8)$~km~s$^{-1}$ \citep{cgk+98} was made by using the white dwarf companion of the pulsar.

\subsubsection{PSR\,J1022+1001}
According to our analysis of the EPTA data, PSR\,J1022+1001 has a $\ddot{\nu}$ in  $(-2.6$,\,$-0.4)\times10^{-28}$\,s$^{-3}$ (95 per cent C.L.). This pulsar has been observed by both the EPTA and PPTA, however, the timing baseline of the EPTA data set is 17.5 years, about twice that of the PPTA. This difference is probably the main reason that the $\ddot{\nu}$ of the EPTA data set has a much narrower PDF than that of the PPTA data. 

 The red noise of this pulsar has a small power spectrum index of 1.6 \citep{cll+16}, suggesting further noticeable decrease of $\sigma(\ddot{\nu})$ and leading to an increased ability to detect  $\ddot{\nu}$ when the timing baseline increases. 

\subsubsection{PSR\,J1024$-$0719}
This pulsar was reported to have significant $\ddot{\nu}\approx-4\times10^{-27}$\,s$^{-3}$, which can be explained by the gravitational jerk caused by a remote companion in a wide binary system \citep{bjs+16,kkn+16}. The reported values are confirmed by our result, $\ddot{\nu}=(-3.93\pm0.46)\times10^{-27}$\,s$^{-3}$, from the PPTA data with high significance (> 8). They are also consistent with the constraint from the EPTA data, see Table~\ref{SigF2}. The consistency between the literature and our results confirms the validity of our method in fitting for $\ddot{\nu}$.

However, the measurement uncertainty reported by \cite{bjs+16}, $\sigma(\ddot{\nu})=2\times10^{-29}$\,s$^{-3}$, is one order of magnitude smaller than what we obtained. We note that the data set used in \cite{bjs+16} spans $\sim22$ years, about 5 years longer than the EPTA data set and 7 years longer than the PPTA data set. More importantly, the timing cadence in the data set of \cite{bjs+16} are much higher than that of the EPTA data, which have only four TOAs in the first 3 years and a long gap in the following 6 years. In addition, \cite{bjs+16} included a DM variation model but no treatment of red noise. Both the better data quality and the lack of red noise analysis in the fitting process are responsible for the small $\sigma(\ddot{\nu})$ reported by \cite{bjs+16}.

\subsubsection{PSR\,B1821$-$24A}
The constraint on $\ddot{\nu}$ for PSR\,B1821$-$24A gives $(29.42\pm15.75)\times10^{-27}$\,s$^{-3}$, which has significance less than 2$\sigma$, while two previous results strongly supported a non-zero value with $\ddot{\nu}=(-175\pm7)\times10^{-27}$\,s$^{-3}$ \citep{cbl+96} and $\ddot{\nu}=(-26.49\pm0.05)\times10^{-27}$\,s$^{-3}$ \citep{jgk+13}. 

PSR\,B1821$-$24A has an usually high apparent $\dot{\nu}$, which may be significantly affected by the potential of its host cluster M28 (NGC\,6626). The pulsar is only $10\arcsec.9$ from the cluster centre, using the cluster position from \citet[2010 edition]{har96} and the pulsar position from \cite{rhc+16}. Using a velocity dispersion of 12.6\,km\,s$^{-1}$, a core size of 0.13\,pc, and a distance of 5.5\,kpc for M28 \citep[table 2]{bh18}, we followed \citet[equation 5]{frk+17} and computed the contribution to $\dot{\nu}$ from the acceleration caused by the cluster potential. This factor contributes at most 15 per cent to the observed $\dot{\nu}$, i.e. the intrinsic $\dot{\nu}$ can be at most 15 per cent larger or smaller. We thus confirmed the statement by \cite{jgk+13} that the apparent spin-down rate is affected only slightly by the cluster potential. 

To explain the origin of the possibly high $\ddot{\nu}$, we estimated the six different contributions to the apparent $\ddot{\nu}$ presented by \citet[equation 6]{lbs18}. Using the apparent $\dot{\nu}$ as the intrinsic spin down, the first contribution is due to radiation  braking and assuming a braking index of 3 gives $\ddot{\nu}=3\dot{\nu}^2/\nu\approx3\times10^{-28}$\,s$^{-3}$. The second contribution, caused by the intrinsic spin-down and the proper motion, depends on the measurement of the currently unknown proper motion. Using the proper motion of M28 from \citet[table 5]{cgj+13} instead, we obtained an estimate of $\sim10^{-31}$\,s$^{-3}$. We set an upper limit of $\sim10^{-29}$\,s$^{-3}$ on the absolute value of the third term (Galactic acceleration term) of equation 6 in \citet{lbs18} by accounting for the acceleration caused by the cluster potential. The last three corrections in the equation are induced by the spatial motion or the Galactic jerk. Using a mean radial velocity of 11.0\,km\,s$^{-1}$ for the cluster and an escape velocity of 49.5\,km\,s$^{-1}$ \citep[table 2]{bh18}, the pulsar radial velocity should be less than 60\,km\,s$^{-1}$ if it is bound to the cluster. We then followed the numerical method in \citet[section 3.2]{lbs18} and found these three terms are on order of $10^{-31}$\,s$^{-3}$ or smaller. They are thus insufficient to explain a $\ddot{\nu}$ as high as $10^{-27}$\,s$^{-3}$.

Two remaining possible origins are the jerk caused by the cluster potential \citep{phi92,phi93,prf+17,frk+17} and that by a passing cluster star \citep{phi92,phi93,frk+17}. Following \citet[equation 10]{frk+17} and using a maximum velocity of 49.5\,km\,s$^{-1}$ \citep[table 2]{bh18}, we find that the upper limit on $|\ddot{\nu}|$ from the cluster jerk is $3\times10^{-25}$\,s$^{-3}$. According to \citet[equation 3.3]{phi92} and \citet[equation 58]{prf+17}, the characteristic contribution from a neighbouring star can be estimated from the local mass density and from the relative velocity between the pulsar and the star. As the pulsar is close to the cluster centre (the angular distance to the cluster centre is only 2.2 times the angular core size), we adopted the core density of $2.6\times10^7$\,M$_\odot$\,pc$^{-3}$  \citep[table 2]{bh18} as the local density. Using the  
escape velocity mentioned before, the maximum characteristic $\ddot{\nu}$ can be produced is $\sim 6.4\times10^{-24}$\,s$^{-3}$. Therefore, both factors can lead to a $\ddot{\nu}$ on order of $10^{-27}$\,s$^{-3}$, depending on the geometry of the system. As both contributions vary with a timescale of $\sim100$ years or longer, it is not likely to observe a significant change in $\ddot{\nu}$ in one to two decades. We suspected the two different values reported by \cite{cbl+96} and \cite{jgk+13} were caused by imperfect treatments of the timing data. Their significantly smaller error bars are probably due to the timing results not considering a noise model.

\subsubsection{PSR\,B1855+09}
Using the EPTA data, we confirmed the conclusion of \cite{ktr94} that no significant $\ddot{\nu}$ is detected in PSR\,B1855+09. Furthermore, by combining the data from \cite{ktr94} and the EPTA, we reduced the uncertainty $\sigma(\ddot{\nu})$, by a factor of $\sim10$, from $9\times10^{-28}$\,s$^{-3}$ to $8\times10^{-29}$\,s$^{-3}$ thus narrowed down the confidence interval with 95 per cent C.L. to $(-2.6$,\,$0.4)\times10^{-28}$\,s$^{-3}$, which is still much larger than the predicted value of $|\ddot{\nu}|\sim10^{-31}$\,s$^{-3}$ that can be caused by radial velocity, Galactic acceleration and jerk \citep[figure 1]{lbs18}. 

\subsubsection{PSR\,J1909$-$3744}
The analysis failed due to a numerical instability when we attempted to obtain the PDF of $\ddot{\nu}$ of PSR~J1909$-$3744 using the timing data and pulsar parameters from the PPTA. We determined that the numerical instability was due to the choice of binary model in the PPTA parameter files. Since the small eccentricity of the orbit is on order of $\sim 10^{-7}$, it is beneficial to use the ELL1 binary model to alleviate the strong correlation between longitude and reference epoch of orbital passage \citep{lcw+01}. To make a successful analysis of PSR~J1909$-$3744, we replaced all the binary parameters with those from the EPTA\footnote{In the timing ephemeris, we still specified the T2 binary model, which will use the ELL1 parameters and include the Kopeikin terms \citep{ehm06}.}. We kept using other pulsar parameters and timing data from the PPTA. The resultant PDF of $\ddot{\nu}$ is consistent with that of the EPTA and both PDFs indicate non-detection of $\ddot{\nu}$. We thus  strongly recommend the use of the extended ELL1 binary model parameterisation \cite[e.g.][]{ehm06, sgj+18} for  PSR~J1909$-$3744 and other highly circular binaries for the purpose of correct timing analysis. This  avoids the risk of numerical instabilities in, for example, constraining the strength of gravitational-wave background.

\subsubsection{PSR\,B1937+21}
This pulsar is very interesting due to the obvious long-timescale structure in the residuals, although the uncertainty of each timing residual is very small \citep{rhc+16,dcl+16,abb2+18}. The similar residual structure was interpreted by \cite{ktr94} as an effect of red noise, while \cite{scm+13} attempted to explain it with an asteroid belt.   

We fitted the timing data from the EPTA and the PPTA to a timing model which included $\ddot{\nu}$. Both data sets consistently give a high $\ddot{\nu}$ of $\sim9\times10^{-27}$\,s$^{-3}$ with a high significance of $\sim4$, see Table~\ref{SigF2}. Using our new measurement uncertainty, the value of $\ddot{\nu}$ is consistent with that obtained by \cite{ktr94} within $3\sigma$, however, our $\sigma(\ddot{\nu})$ is $\sim5$ times larger than that in \cite{ktr94}. This increase of uncertainty could be caused by our inclusion of DM variations and red noise in our analysis.  

To see if $\sigma(\ddot{\nu})$ can be reduced when the timing baseline is longer, we did a further fitting for $\ddot{\nu}$ by combining the data from \cite{ktr94} and the EPTA, extending the baseline from 24.1 years to 29.5 years. A $\ddot{\nu}$ and significance consistent with those using EPTA or PPTA data separately were recovered, while $\sigma(\ddot{\nu})$ surprisingly increased.

The unusual increase of $\sigma(\ddot{\nu})$ contradicts the general expectation that the measurement error should decrease with longer data sets (see Section~\ref{scaling_expression}). The reason is still unknown. 

Since the independent constraint from the PPTA, the EPTA and the combined data gave consistent $\ddot{\nu}$ and $\sigma(\ddot{\nu})$, the $\ddot{\nu}$ of $\sim9\times10^{-27}$\,s$^{-3}$ is very possibly a real signal, although a physical explanation is required to support or verify this idea. A remote companion with an orbital period much larger than the current timing span may be responsible for such a $\ddot{\nu}$, as in the case of PSR\,J1024$-$0719 \citep{bjs+16,kkn+16}. To investigate this hypothesis, we firstly obtained the frequency derivatives by fitting the timing data of PSR~B1937+21, with a polynomial series of spin frequency derivatives up to $\nu^{(8)}$ to capture both red noise and the real frequency derivatives. The DM variations were also included with the parameters obtained by \textsc{enterprise} analysis and fitted in \textsc{tempo2}. As a result, derivatives with high significance up to $\nu^{(5)}$ were obtained. We then followed the method of \citet[section 4.5]{bjs+16} and used the expression in Appendix~\ref{appen} to find the possible orbital parameters that give bound (circular or elliptic) orbits and satisfy the five frequency derivatives. As the dynamical contribution to $\dot{\nu}$ from the presumed binary component is unknown, we assumed it to be a fraction of the apparent $\dot{\nu}$ with the fraction in the interval of $[-1, 1]$. No suitable solutions were found for these cases. We thus conclude that a remote bound companion can not explain the value of $\ddot{\nu}$.

\begin{figure}
 \includegraphics[width=8.5cm]{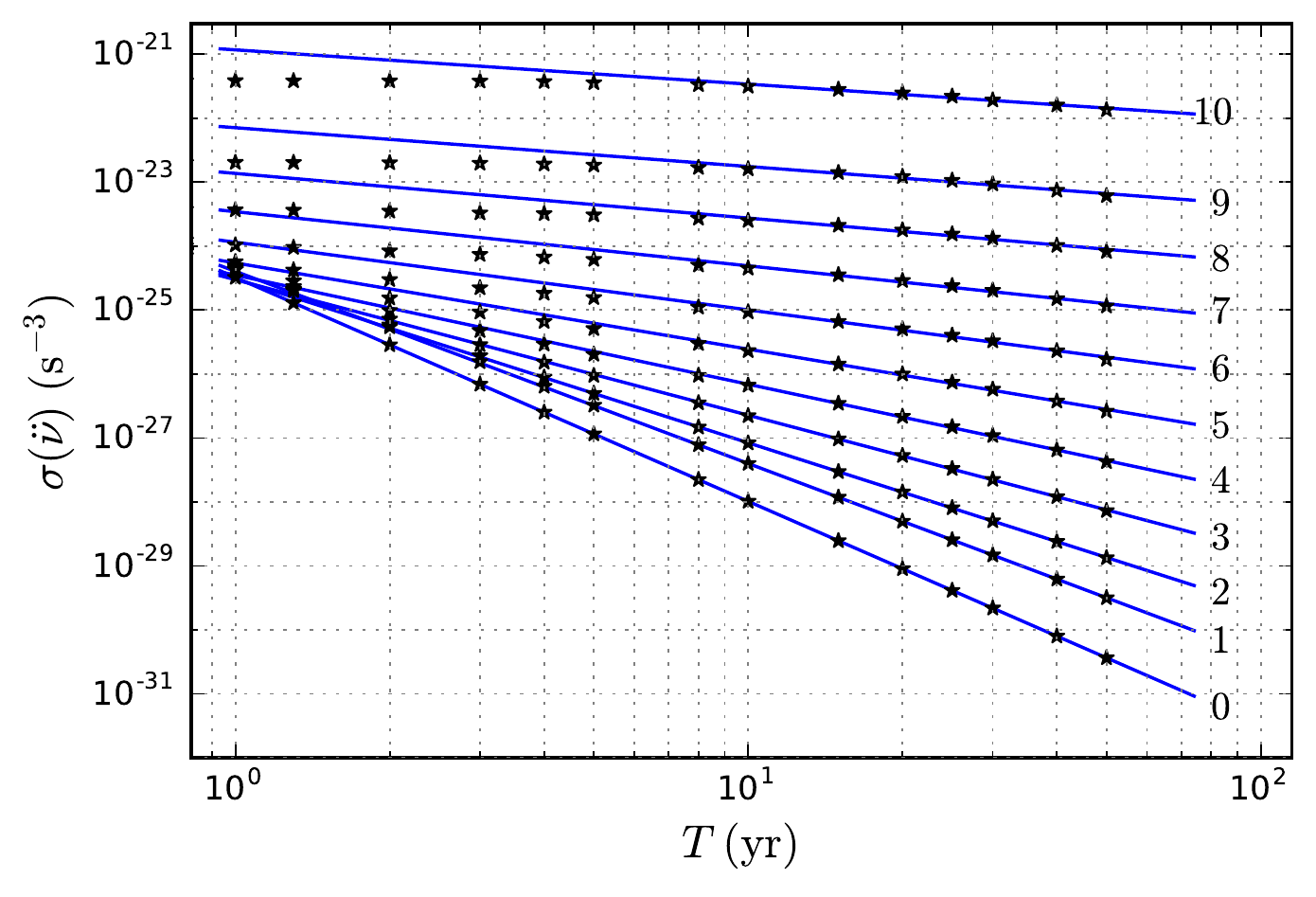}
  \caption{The measurement uncertainty obtained from simulations (black stars) and the best linear fit (blue lines) for $T\ge10$ years, assuming a power law spectrum of red noise. From the bottom to the top, the lines are for $\alpha=0$, 1, 2, $\cdots$, $10$. Points in the case of $\alpha=0$ are computed from \citet[equation 7]{lbs18} by using the same interval and uncertainty as those of other $\alpha$.}
  \label{Red_SigF2}
\end{figure}

\begin{figure}
 \includegraphics[width=8.5cm]{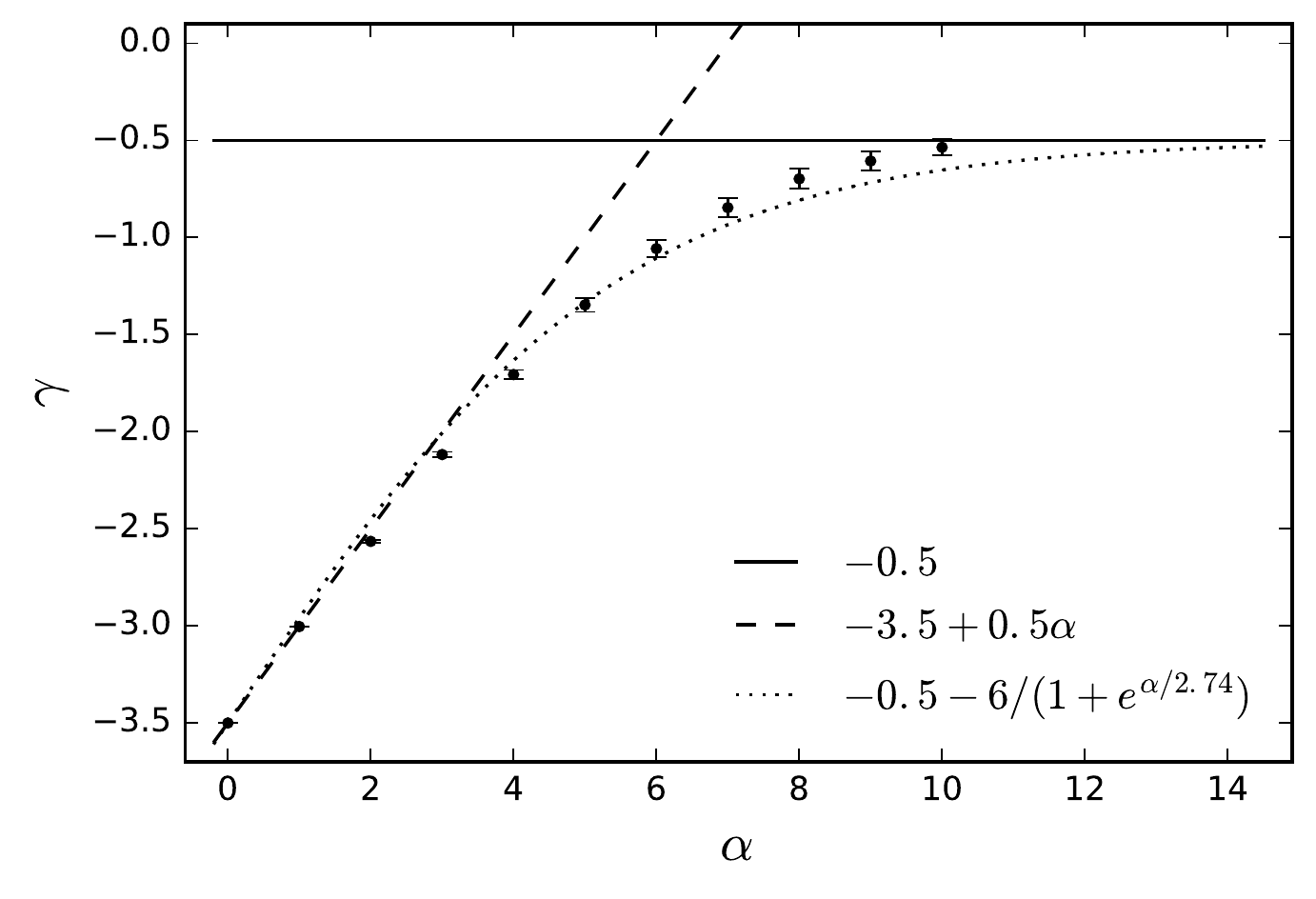}
  \caption{The relation between coefficient $\gamma$ and the index of power spectrum  $\alpha$. The points with an error bar show the slope obtained from the linear fit in Fig.~\ref{Red_SigF2}. The dashed line is the result of a linear fit to the points for $0\le \alpha \le 4$, while the dotted line is for a global non-linear fit to all data points.}
   \label{Scale_law}
\end{figure}

\subsection{Scaling relation of \texorpdfstring{$\sigma(\ddot{\nu})$}{nu}}
\label{scaling_expression}
The relation between $\sigma(\ddot{\nu})$ and the timing baseline is important when assessing the possibility of measuring $\ddot{\nu}$. \cite{lbs18} gave the analytical expression for the case of white noise. Here, we include red noise and re-compute $\sigma(\ddot{\nu})$ by means of simulation. 

We firstly simulated TOAs with an equally separated sampling of 14\,days and a white noise level of $\sigma_{\rm w}=1\,\mu$s. The timing baseline, $T$, was allowed to span from 1 to 50 years and the reference epoch for the spin parameters was chosen to be the mid point. We then modelled the red noise using a power law spectrum in Eqn.~\ref{RedPL}. In the power law, we set $f_{\rm c}$ to be 1/(100\,yr), where 100 yr was chosen to be much larger than the timing baseline. The spectrum is pivoted at a frequency $f_{\rm yr}=1$\,yr$^{-1}$ with the amplitude of the white noise, i.e. $P(f_{\rm yr})=P_{\rm w}$, where $P_{\rm w}$ is the power spectral density of the white noise. The spectral index was allowed to vary from 0 (white noise) to 10, which is sufficient to cover the reported values of the power index of red noise \citep{rhc+16,cll+16,abb2+18}. Finally, we followed the semi-analytical method outlined in Appendix~\ref{compute_sigma_nu2} to obtain $\sigma(\ddot{\nu})$. 

Fig.~\ref{Red_SigF2} shows $\sigma(\ddot{\nu})$ with respect to different values of $T$. The data corresponding to each value of $\alpha$ were fitted with a relation of the form $\log_{10}(\sigma(\ddot{\nu})/{\rm s}^{-3})=\gamma\log_{10}(T/{\rm yr})+b$, where $\gamma$ and $b$ are determined by fitting the data in Fig.~\ref{Red_SigF2} and $\gamma$ indicates the speed with which $\sigma(\ddot{\nu})$ decreases with $T$. We then fitted the $\gamma$ value of each $\alpha$ in Fig.~\ref{Scale_law} and obtained the scaling law for $\alpha\in[0, \infty)$ 
\begin{equation}
\label{scaling}
\Big(\frac{\sigma(\ddot{\nu})}{{\rm s}^{-3}}\Big)=K\Big(\frac{T}{\rm yr}\Big)^\gamma \quad {\rm with} \quad \gamma=\frac{-6}{1+e^{\alpha/2.74}}-\frac{1}{2},
\end{equation}
where $K$ is a function of spin frequency, observation cadence, white noise level, the amplitude and the corner frequency of the red noise. When $0\le\alpha\le4$, $\gamma$ is linearly proportional to $\alpha$ with $\gamma=-3.5+0.5\alpha$. In the limit of very steep red noise, Fig.~\ref{Red_SigF2} suggests different trends on short and long timing baselines. For short timing baselines, where $T$ is much smaller than the correlation length scale of the red noise, $\sigma(\ddot{\nu})$ is dominated by correlated noise on timescales much longer than the timing baseline, therefore does not vary significantly with $T$. On timing baselines where $T$ is larger than the red noise correlation length scale, the number of independent estimates of $\ddot{\nu}$ increases linearly, but slowly, with time, so $\sigma(\ddot{\nu})$ decreases as $T^{-1/2}$. According to Eqn.~\ref{scaling}, for the same observational settings of a particular pulsar, the red noise makes $\sigma(\ddot{\nu})$ increase rapidly with $\alpha$, compared to the case of white noise. Using this relation, we can predict $\sigma(\ddot{\nu})$ for a longer timing baseline, if the other observational factors, like the cadence and the white noise level, remain unchanged.

\subsection{Time to detect \texorpdfstring{$\ddot{\nu}$}{nu} induced by radial velocity}
\label{F2Vr_baseline}

The measurement of $\ddot{\nu}$ due to the radial velocity of a pulsar and the measurement of the velocity itself were considered by \cite{lbs18}. Here we update the predictions for measuring $\ddot{\nu}_\parallel$, the $\ddot{\nu}$ induced by the radial velocity, by considering the effects of red noise. According to Eqn.~\ref{scaling}, the time span to achieve a measurement uncertainty of $\sigma(\ddot{\nu})$ is
\begin{equation}
T=T_0\bigg(\frac{\sigma(\ddot{\nu})}{\sigma_0(\ddot{\nu})}\bigg)^{1/\gamma},
\end{equation}
where $\sigma_0(\ddot{\nu})$ and $T_0$ are the current uncertainty and timing baseline. For a 5$\sigma$ detection, $\sigma(\ddot{\nu})=|\ddot{\nu}_\parallel|/5$. The value of $\ddot{\nu}_\parallel$ depends on the radial velocity. Only four of the 49 MSPs have a measured radial velocity (Table~\ref{F2ms}). For these four MSPs, we used the measurements to compute the corresponding $\ddot{\nu}_\parallel$, while for the remaining pulsars we used an assumed and adequate value of 50\,km\,s$^{-1}$. The value of power index of red noise was taken from \cite{cll+16} or \cite{rhc+16}. In the case of no reported value, we considered two special cases: no red noise, i.e. $\alpha=0$ and a typical power  spectrum index of red noise with $\alpha=3$.

Using the uncertainty of $\ddot{\nu}$ and the current time span in Table~\ref{F2ms}, we list the predicted timing baseline required to make a significant measurement in the columns labelled with $T_1$ and $T_2$. The value of $T_1$ corresponds to the case where only white noise is important, while $T_2$ is calculated using the $\alpha$ listed in the table. A white noise only value is calculated whenever there is no previously published $\alpha$. In both cases, the shortest timing baseline required is on the order of 100 years, thus for this sample of pulsars a detection is unlikely in the near future.     

\subsection{$\ddot{\nu}$ and gravitational wave detection}

According to the detection status of $\ddot{\nu}$, the MSPs can be divided into three categories. In the first category, an MSP, like PSR\,J1024$-$0719, has a confirmed $\ddot{\nu}$, which can increase the timing residuals in a cubic pattern. Modeling the confirmed $\ddot{\nu}$ in the pulsar ephemeris is helpful to reduce the rms residuals, improve its timing precision and minimize the impact on signals from gravitational waves. The second category contains the MSPs that are on the verge of $\ddot{\nu}$ detection (or non-detection), including PSRs\,J0621+1002, J1022+1001, B1821$-$24A and B1937+21. For these MSPs, the current imprint of $\ddot{\nu}$ is not clearly distinguishable from that of red noise. Extending the timing baseline may reduce $\sigma(\ddot{\nu})$ and constrain $\ddot{\nu}$ better. In addition, investigating potential sources that can cause high $\ddot{\nu}$, such as unidentified remote companions, and computing the corresponding range of such $\ddot{\nu}$ are helpful to accept or reject a $\ddot{\nu}$. In the third category (most pulsars in the IPTA are in this category), the $\ddot{\nu}$ is currently undetectable due to the very small $\ddot{\nu}$ as predicted by the theory \citep{lbs18}. The rms residuals caused by $\ddot{\nu}$ are thus very small and the resultant impact on gravitational waves can be neglected.

\section{Conclusions}
\label{conclusions}
We have searched for the unmodelled $\ddot{\nu}$ in a pulsar ephemeris by using a Bayesian approach. We included models of red noise and DM variations and adopted \textsc{enterprise} to efficiently sample the possible parameter spaces. Our method was validated by the successful recovery of $\ddot{\nu}$ from the simulations. The robustness of the method was also tested by fitting for the red noise of squared exponential kernel with a power law model, although tests using more types of red noise model are necessary to obtain general conclusions. We further note that the methodology described in this paper provides an approach that could be used more generally when undertaking studies to determine whether additional timing parameters need to be included in the timing analysis. 

After searching the timing data of 49 millisecond pulsars in the EPTA and PPTA, we obtained the marginalised probability density function of $\ddot{\nu}$ (Fig.~\ref{F2pdf}). We thus confirmed the detection of $\ddot{\nu}$ in PSR~J1024$-0719$ and found a statistically significant $\ddot{\nu}$ in PSR~B1937+21. Neither the spin-down due to the  braking process nor a remote binary companion can explain the $\ddot{\nu}$ of PSR~B1937+21.

By computing the measurement error of $\ddot{\nu}$ for power law red noise with different power indices of $\alpha$, we found that the error increases rapidly by $\sigma(\ddot{\nu}) \propto T^\gamma$, where $T$ is the timing baseline and $\gamma=-6/[1+\exp{(\alpha/2.74})]-1/2$. Our results thus quantitatively support the idea that red noise plays an important role in error estimation. Using these error estimates, we further predicted the timescale to measure the $\ddot{\nu}$ caused by the radial velocities of pulsars. For the pulsars we considered, it needs a timing baseline of $\sim 100$ years or longer to detect the $\ddot{\nu}$ or the radial velocity, significantly longer than in the case of white noise.

Our research has revealed that $\ddot{\nu}$ does not generally make a significant contribution to the arrival times (if there is any) of the MSPs observed by the EPTA and the PPTA. However, a few MSPs (e.g. PSR~B1937$+$21) have statistically significant $\ddot{\nu}$ which require physical explanations to confirm or interpret the results. There is also a group of MSPs, like PSRs\,J0621+1002, J1022+1001 and  B1821$-$24A, exhibiting mildly significant $\ddot{\nu}$, which may become more significant as the timing baseline and accuracy increase. For these pulsars the inclusion of $\ddot{\nu}$ in the timing solutions should be revisited.

\section*{Acknowledgements}
XJL acknowledges support from the President's Doctoral Scholar Award from the University of Manchester. XJL would like to thank Benetge Perera, Benjamin Shaw and Thomas Scragg for useful discussions. We appreciate the generous supply of computational resource from Rene Breton. Pulsar research at Jodrell Bank Centre for Astrophysics is supported by a Consolidated Grant (ST/P000649/1) from the UK's Science and Technology Facilities Council.

\appendix
\section{The $q_6$ expression}
\label{appen}
Spin frequency derivatives up to the fourth order were neatly given by \citet{bjs+16}. Here we present the expression of the fifth derivative. Using the same conventions in the aforementioned reference, we have $f^{(5)}=-fz^{(6)}_1/c$, where 
\begin{equation*}
    z_1^{(6)}=\frac{k^3\sin{i}}{a_1^8}q_6(e, \nu, \omega),
\end{equation*}
and 
\begin{equation*}
 \begin{split}
    q_6=&\frac{1}{16}\frac{(1+e\cos{\nu})^6}{(1-e^2)^8}\Big[-40e^2\sin{(\omega-\nu)}+180e^4\sin{(\omega-\nu)}\\
    & -105e^4\sin{(\omega-3\nu)}+32e\sin{\omega}-312e^3\sin{\omega}-16\sin{(\omega+\nu)}\\
    &+496e^2\sin{(\omega+\nu)}-270e^4\sin{(\omega+\nu)}-448e\sin{(\omega+2\nu)}\\
    &+1008e^3\sin{(\omega+2\nu)}-1960e^2\sin{(\omega+3\nu)}+420e^4\sin{(\omega+3\nu)}\\
    &-2520e^3\sin{(\omega+4\nu)}-945e^4\sin{(\omega+5\nu)} \Big].
     \end{split}
\end{equation*}

\section{Computing $\sigma(\ddot{\nu})$}
\label{compute_sigma_nu2}

The measurement uncertainty $\sigma(\ddot{\nu})$ can be estimated by a generalized least-squares fit to $\mathbf{y}=\mathbf{X}\beta$, where $\mathbf{y}=\{\phi_1, \phi_2,\,...\,,\phi_N \}^\intercal$ includes the pulse phase of $N$ observations, $\mathbf{X}$ is design matrix with elements of $\mathbf{X}_{ij}=t_i^{j-1}$ ($t_i$ is the $i-$th TOA with $i$ ranging from 1 to $N$, while $j$ from 1 to 4) and $\beta=\{1, \nu, \dot{\nu}/2, \ddot{\nu}/6\}^\intercal$ contains the spin parameters.

The covariance matrix is ${\rm cov}(\mathbf{\beta})=(\mathbf{X}^\intercal\mathbf{W}\mathbf{X})^{-1}$, where $\mathbf{W}$ is weight. The weight is the matrix inverse of the covariance of red noise, or $\mathbf{W}=\mathbf{C}^{-1}$. To make the computations efficient and stable, we decomposed $\mathbf{C}$ by the lower triangular Cholesky factorization through $\mathbf{C}=\mathbf{L}\mathbf{L}^\intercal$ thus ${\rm cov}(\beta)=\big((\mathbf{L}^{-1}\mathbf{X})^\intercal(\mathbf{L}^{-1}\mathbf{X})\big)^{-1}$. A further QR decomposition (e.g. \citealt{ptv+02}), $\mathbf{L}^{-1}\mathbf{X}=\mathbf{Q}\mathbf{R}$, gave us the final expression of the covariance matrix, ${\rm cov}(\beta)=\mathbf{R}^{-1}(\mathbf{R}^{-1})^\intercal$. 

We obtained the covariance function of the red noise by the \texttt{analyticChol} plug-in in \textsc{tempo2} \citep{hem06} and interpolated it to generate $\mathbf{C}$. The red noise input has been described in the main text (Section~\ref{scaling_expression}) and was pivoted at the white noise. The spectral density of white noise can be expressed in terms of the time span and the number of TOAs by $P_{\rm w}=2T\sigma_{\rm w}^2/N$ \citep[section 2]{kcs+13}, when the TOAs have an equal uncertainty of $\sigma_{\rm w}$. 

\bsp	
\label{lastpage}
\bibliographystyle{mnras}
\bibliography{F2Noise.bib}
\end{document}